\documentclass{article}

\usepackage{arxiv}

\usepackage[utf8]{inputenc} 
\usepackage[T1]{fontenc}    
\usepackage{hyperref}       
\usepackage{url}            
\usepackage{booktabs}       
\usepackage{amsfonts}       
\usepackage{nicefrac}       
\usepackage{microtype}      
\usepackage{lipsum}
\usepackage{graphicx}
\usepackage{comment}
\usepackage{xcolor}
\usepackage{amsmath}
\graphicspath{ {./images/} }

\title{
A job-based assessment of economic complexity: from hidden to revealed
}

\author{
 Antonio Russo \\
  Department of Economics and Finance, \\ University of Rome Tor Vergata, Italy
   \\
   \And
 Pasquale Scaramozzino \\
  Department of Economics and Finance, \\ University of Rome Tor Vergata, Italy, and \\ School of Finance and Management, \\ SOAS University of London, U.K.\\
   \\
  \And
 Andrea Zaccaria \\
  Institute for Complex Systems\\
  National Research Council, Italy\\
  \texttt{andrea.zaccaria@cnr.it} \\
}

\begin{document}
\maketitle
\begin{abstract}
Economic complexity measures aim to quantify the capability content or endowment of industries and territories; however, capabilities are not observable, and therefore cannot be directly used in the computations. We estimate such endowments by quantifying the quality and diversity of the skills in the occupations required in specific industries. We refer to this job-based assessment as the hidden complexity, in contrast with the usual revealed complexity which is computed from economic outputs such as exports or production. We show that our job-based measure of complexity is positively associated to wage levels and labor productivity growth, whereas the classic revealed measure is not. Finally, we discuss the application of these methods at the territorial level, showing their connection with economic growth.
\end{abstract}


\section{Introduction}

Economic complexity \cite{balland2022reprint,hausmann2014atlas,hidalgo2021economic,diodato2024handbook} is a data-driven approach that adopts tools such as network science and machine learning to investigate development economics and the economics of innovation. Despite its widespread use in both prediction and policy analyses, it lacks a solid theoretical foundation \footnote{Notable exceptions are the recent attempts to mathematically derive the equations of the economic complexity algorithms from optimization principles: see for instance \cite{mazzilli2024equivalence,mariani2024ranking}.}.  The baseline for the analysis is usually a bipartite network that connects an economic actor or territory (such as a country or a firm) to an economic activity (such as exporting a product or patenting in a specific technology field) \cite{hidalgo2009building,tacchella2012new}. Then, one algorithmically computes an assessment of the economic complexity of these nodes, where complexity is intended to quantify the capability content of an activity or the presence of such capabilities in a country, region, or firm. However, while one  may have some intuition of what might be a sophisticated product or a developed country, it is very hard to define these “capabilities” in a precise way. Indeed, the meaning of capability in the context of firms is completely different, so that one usually speaks about the “organizational capabilities” of companies \cite{dosi2000nature,costa2023organizational}, while in the economic complexity field capabilities are “building blocks” \cite{hidalgo2009building,tacchella2012new,zaccaria2014taxonomy,fagerberg2010innovation} that, in some sense, are required for a country to be competitive in a specific industry - often without even making examples about what a capability might be. Conceptually, one builds a \textit{tripartite} network connecting countries, capabilities, and products, and argues that a country can competitively export a product if it is connected to all the capabilities needed to export that product \cite{hidalgo2009building,tacchella2012new,zaccaria2014taxonomy,balland2022reprint}. The economic complexity measures are built starting from this conceptual framework. In particular, a product is said to be “complex” if it needs many or rare capabilities, and analogously, an economy is complex if it has many or rare capabilities (note that other operative definitions of complexity are possible, see for instance \cite{broekel2019using}). The problem of translating this tripartite theoretical framework into empirical data is not only the availability of cross-country harmonized capability data \cite{fagerberg2010innovation}, but also the high dimensionality of the relative simplicity of the model with respect to the high dimensionality of the system and of its interactions \cite{de2025bayesian,tian2025matrix}. The economic complexity approach infers the capability content by looking at the revealed comparative advantage  \cite{hausmann2007you,hidalgo2009building,tacchella2012new}: one extracts this information from the visible bipartite actor-activity network because the hidden layer is practically unobservable. Indeed, while products, patents, and even scientific activities are available in databases that are internationally harmonized \cite{pugliese2019unfolding}, this is not true for capabilities, for which an agreed-upon definition is lacking. For instance, in \cite{cristelli2018virtuous} the authors connect the development stage of nations with the presence of infrastructures such as railways and internet availability, while Vu \cite{vu2022does} focuses on institutional quality. 
In this paper we focus on human capital \cite{zhu2017economic,sadeghi2020economic,osinubi2025economic,sbardella2018role,hosseinioun2025skill} as the essential element of the capability structure of production, and derive new assessments of the economic complexity of industries and territories starting from occupational data. Some attempts in this direction are present in the literature. Schetter et al. \cite{schetter2024products} and \cite{neffke2013skill} build measures of the relatedness between industries, leveraging human capabilities as proxied by the occupations recorded in each sector. This allows Schetter et al. to build what they call a \textit{genotypic} product space, that is, industry-to-industry connections based not on the (phenotypic) output, as is usually done in economic complexity \cite{hidalgo2007product,zaccaria2014taxonomy,albora2022machine}, but directly on the human capabilities \cite{neffke2013skill}. Lo Turco and Maggioni \cite{turco2022knowledge} define a measure of occupational complexity by looking at the skill content of industries, and show that it can predict the economic growth of cities. Buyukyazici et al. \cite{buyukyazici2024workplace} introduce a measure of the skill complexity of industries by applying the EC algorithms to the skill-industry network.
The present paper agrees with this line of thought, which recognizes human capital as a key determinant for the economic complexity of industries and territories. Our approach, however, differs from previous investigations in both the empirical strategy and the conceptual discussion of the results. First of all, we leverage the results obtained by Aufiero et al. \cite{aufiero2024mapping} 
to quantify the \textit{Fitness of Jobs}, an assessment of the sophistication of occupations which is computed by considering a complexity-weighted diversification of the skills the jobs require. By aggregating and averaging the Job Fitness values, we can compute (human) capability-based assessments of the complexity content of territories and industries. This \textit{hidden complexity} can then be compared with the \textit{revealed complexity}, following the genotypic vs phenotypic argument of \cite{schetter2024products}, where the revealed quantities are computed in the standard way, that is, by looking ex-post at the effective output (in this case, the export of countries). This comparison scheme offers a number of insights about the productive structure and its relation with human capital. Finally, we show how our job-based measure of economic complexity is positively associated with local economic growth and, at the industry level, with the growth of labour productivity - a result which could not be obtained using the standard, export-based complexity measure.

\section{Data description and preprocessing}

\begin{figure}
\centering
  \includegraphics[width=\textwidth]{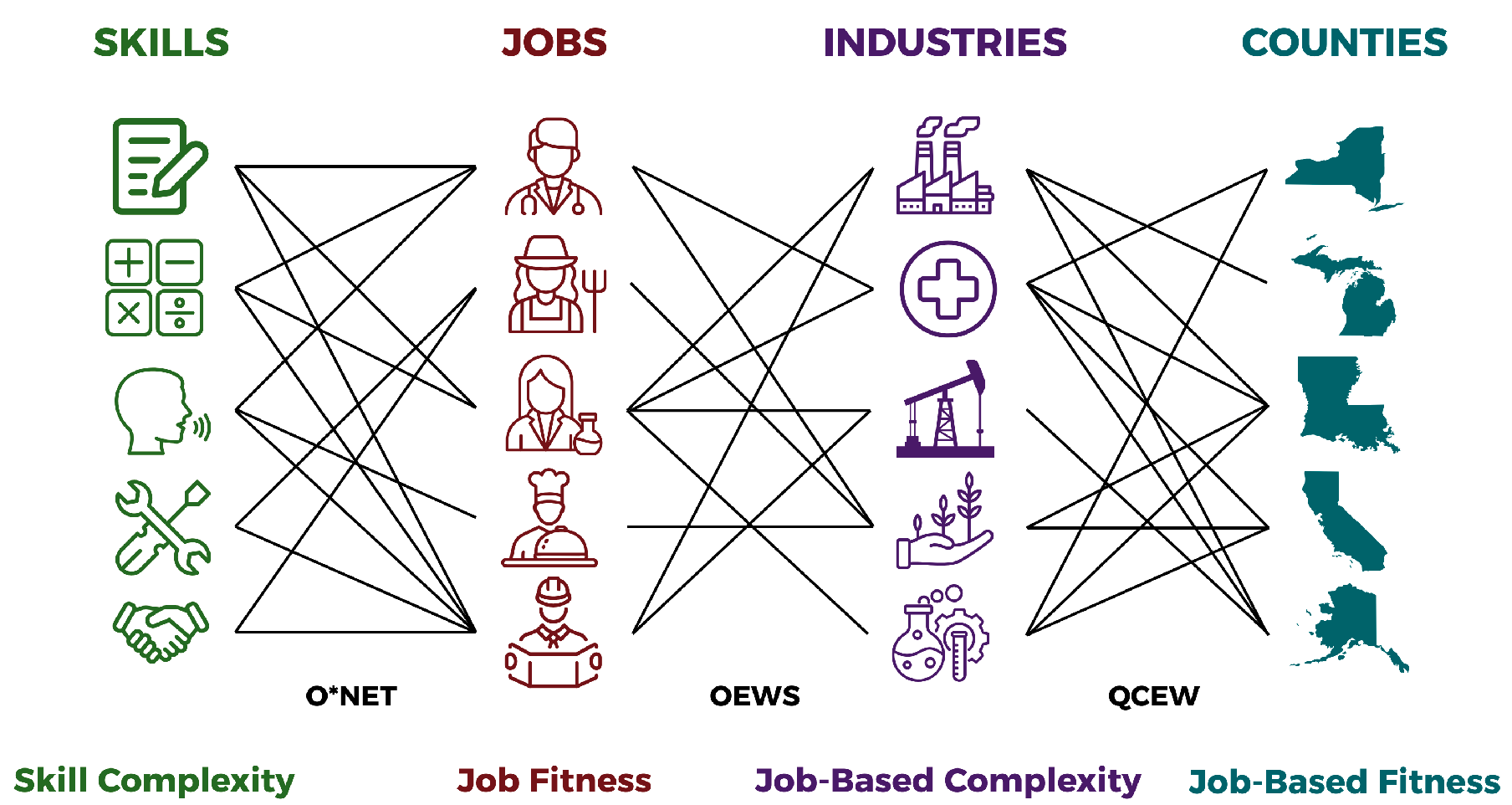}
  \caption{An illustration of the starting databases (in black) and the measures we compute (colored writings at the bottom). Our framework can be represented as a quadripartite network, with four sets of nodes whose connections are established using three databases. From left to right, we apply the Fitness algorithm to determine the Skill Complexities and the Job Fitnesses, which we average to compute the Job-Based Complexity of industries that we sum to assess the Job-Based Fitness of counties.}
  \label{fig:4part}    
\end{figure}

In Figure \ref{fig:4part} we schematically represent our multipartite approach. We consider four different layers of information: skills, jobs, industries, and US counties. These four layers can be represented as nodes connected via three bipartite networks: from left to right, the skill-job bipartite network, extracted from the O*NET database; the job-industry network, from the BLS database; and the industry-county network, also from the BLS database. In the following, we will indicate the adjacency matrices which define these networks by $\mathbf{M^{(1)}}$, $\mathbf{M^{(2)}}$, and $\mathbf{M^{(3)}}$, respectively. Finally, we use the UN-COMTRADE database to compute the standard, “revealed” economic complexity of industries and counties.

\subsection{O*NET database to connect skills and jobs}

The O*NET database  (www.onetonline.org) is a comprehensive resource for occupational information, detailing job requirements, skills, and worker characteristics. Following \cite{aufiero2024mapping}, we use it to build the bipartite network $\mathbf{M^{(1)}}$, which connects 68 skills with 439 occupations. In particular, O*NET provides measures of the \textit{importance} of the specific skill for each occupation using a scale from 1 to 5. The occupations are classified according to the SOC (Standard Occupational Classification). In order to obtain a binary network, we keep only those links whose weights exceed the average importance of that skill (considering all occupations).

\subsection{BLS database to connect jobs with industries and industries with counties}

The Bureau of Labor Statistics (BLS) (www.bls.gov) provides employment and wage estimates across industries and US geographic areas. In this paper, we focus on the Occupational Employment and Wage Statistics (OEWS) and the Quarterly Census of Employment and Wages (QCEW) databases to establish the connections between Jobs and Industries and between Industries and Counties, respectively (see Figure \ref{fig:4part}). In particular, we consider the wage bills associated with each one of the SOC occupations (435 different occupations, since not all occupations are available in this data set), and for which we also know the industry information. Industries surveyed by OEWS are classified according to the 4-digit NAICS classification, which contains 250 different nonfarm industries. However, 30 of these industries cannot be matched with the macroeconomic data used for regression analysis. Thus, in what follows (and specifically when referring to regression analysis) we use 220 industries: 74 are goods-producing (belonging to \textit{Mining, Quarrying, and Oil and Gas Extraction} and \textit{Manufacturing} sectors) and 146 refer to services. In order to build the bipartite Job-Industry network, we first consider the wage bill ${W}_{ji}$ per industry and per occupation and normalize it to take into account both the size of the industry and the total wage associated with the occupation. In analogy with the Location Quotient introduced in \cite{isserman1977location}, we define the Industry Wage Quotient
\[
\text{IWQ}_{ji} = 
    \frac{W_{ji}}{\sum\limits_{j'} W_{j'i}}
\bigg/
    \frac{\sum\limits_{i'} W_{ji'}}{\sum\limits_{j'i'} W_{j'i'}}
\]
Such normalization, in analogy with the use of Balassa's Revealed Comparative Advantage \cite{balassa1965tariff} in Economic Complexity, allows for a natural threshold equal to 1. The Job-Industry bipartite network is represented by the matrix $\mathbf{M^{(2)}}$ whose elements are equal to one if $IWQ_{ji}>1$, and zero otherwise.\\
We apply the same line of reasoning to build the Industry-County network $\mathbf{M^{(3)}}$, which contains 3159 counties classified by FIPS codes. Following \cite{sbardella2017economic}, we define the Wage Location Quotient in analogy with the Location Quotient \cite{isserman1977location}, but considering the wages: 
\[
\text{WLQ}_{ci} = 
    \frac{W_{ci}}{\sum\limits_{c'} W_{c'i}}
\bigg/
    \frac{\sum\limits_{i'} W_{ci'}}{\sum\limits_{c'i'} W_{c'i'}}
\]
where ${W}_{ci}$ is the wage bill in county $c$ for industry $i$. The binary industry-county network is described by the adjacency matrix $\mathbf{M^{(3)}}$ whose elements are equal to one if $WLQ_{ci}>1$, and zero otherwise. Both the Job-Industry and the Industry-County networks are computed using data relative to 2017 and 2022. We use the 2017 data to predict the growth of the macroeconomic quantities in the following years.

\subsection{UN-COMTRADE to compute the exogenous fitness}

The UN-COMTRADE database (www.comtrade.un.org) provides information about international trade, retrieved by local customs and harmonized using a suitable shared classification. This dataset is usually organized in a bipartite network that connects countries to the products they export. More precisely, we compute the Revealed Comparative Advantage \cite{balassa1965tariff} for country $c$ in product $p$ as:

\[
\text{RCA}_{cp} =
    \dfrac{E_{cp}}{\sum\limits_{p'} E_{cp'}}
    \bigg/
    \dfrac{\sum\limits_{c'} E_{c'p}}{\sum\limits_{c'p'} E_{c'p'}}
\]

\noindent where $E_{cp}$ is the exported value, expressed in US dollars, of country $c$ in product $p$. The bipartite country-product network is binarized by imposing $\text{RCA}_{cp} > 1$; so $M_{cp}^{(exp)}=1$ if country $c$ has a comparative advantage in product $p$. This procedure is repeated for all years between 2012 and 2021 (included). This allows us to average the economic complexity quantities derived from the single-year country-product networks.

\subsection{Macroeconomic data}

In order to validate our economic complexity measures, we use macroeconomic data retrieved from various sources. Raw data used to construct measures of average total compensation and labor productivity growth are obtained from Economic Census --- the official five-year survey of U.S. businesses conducted by the U.S. Census Bureau (https://www.census.gov/en.html) --- which provides comprehensive statistics at the national, state, and local levels. Data on revenues, compensation, and number of workers are available at the industry level for most nonfarm industries. The Producer Price Index (PPI) program, published by the Bureau of Labor Statistics (BLS) (https://www.bls.gov/ppi/), measures the average change over time in the selling prices received by domestic producers for their output. We use this index to adjust nominal output growth for price changes over time. Finally, to study local economic growth and its relationship with our economic complexity measures, we use county-level Gross Domestic Product (GDP) estimates released annually by the Bureau of Economic Analysis (BEA) (https://www.bea.gov). All growth rates refer to the 2017-2022 time interval.

\section{Methods}

\subsection{The Fitness of Jobs}

Our assessment of the hidden, capabilities-based complexity of industries requires a quantification of the skill content of the occupations required by that industry. So our first step is to compute the \textit{Fitness of jobs}, an algorithmic assessment of the quality and the quantity of the skills needed by each occupation. Our operational approach follows the economic complexity literature \cite{hidalgo2009building,tacchella2012new}, and in particular \cite{aufiero2024mapping}. The idea is to leverage the bipartite skill-job network identified by the adjacency matrix $\mathbf{M^{(1)}}$, whose element $M^{(1)}_{sj}$ is equal to 1 if skill s is required for job j, and zero otherwise. To quantify the economic complexity, or fitness of job j, we adopt the Economic Fitness and Complexity algorithm introduced by \cite{tacchella2012new}. The idea is to compute the fitness of jobs as a complexity-weighted diversification of skills, and that the complexities of skills are a nonlinear function of the fitness of jobs - in particular, the complexity of a skill is low if at least one low-fitness job requires it. In formulas, we have 
\begin{equation}
F_{j}^{(n)}=\sum_{s} M^{(1)}_{sj} Q_{s}^{(n-1)} \qquad
Q_{s}^{(n)}=\frac{1}{\sum_{j} M^{(1)}_{sj} \frac{1}{F_{j}^{(n-1)}}}
\label{eq:FJobs}
\end{equation}
where $F_j$ is the fitness of job $j$, $Q_s$ is the complexity of skill $s$, and $n$ is the algorithm iteration number. Following \cite{tacchella2012new}, at each iteration we normalize the fitness and the complexity values using the respective mean. The fixed point of this map does not depend on the initial conditions $F_{j}^{(0)}$ and $Q_{s}^{(0)}$ \cite{cristelli2013measuring}. We have iterated the algorithm until convergence is achieved following the criteria discussed in \cite{pugliese2016convergence}. \\ In summary, high-fitness jobs are characterized by a diverse set of complex skills - those skills that only complex jobs require - while low-fitness jobs feature a small number of relatively simple and common skills.

\subsection{Job-based economic complexity}

Using the fitness of jobs, we can compute the hidden, or Job-Based, economic complexity of industries by leveraging the information about the job-industry network defined in the previous sections. In particular, we define $Q^{JB}_{i}$, the Job-Based Complexity of industry $i$, as the weighted average of the fitness of the jobs present in such industry:
\begin{equation}
Q^{JB}_{i}=\frac{\sum_{j} N^{(2)}_{ji}F_j}{\sum_{j} N^{(2)}_{ji}}
\label{eq:compJB}
\end{equation}
where $N^{(2)}_{ji}$ is the number of employees with job $j$ in industry $i$. In this way, we can assess the complexity of industries in terms of their capabilities: for this reason, we talk about a \textit{hidden} complexity, in contrast to the revealed complexity, which is computed using observable economic outputs. In our approach, a low hidden complexity is associated with those industries which are characterized by low-fitness jobs - which in turn imply that a small number of simple skills are needed. By contrast, a high value of hidden complexity shows that the industry under investigation uses occupations that feature many complex skills.\\
Once the Job-Based Complexity of industries is known, we compute the Job-Based Fitness of county $c$ by summing the complexities of the industries present in $c$ (in the Wage Location Quotient sense):
\begin{equation}
F^{JB}_{c}=\sum_{i} M^{(3)}_{ic}Q^{JB}_{i}
\label{eq:fitJB}
\end{equation}
Fitness values are then normalized using the mean across counties.
This will be our evaluation of the economic complexity of territories from the point of view of the local human capabilities. In this sense, this is a \textit{hidden} fitness, in contrast with the \textit{revealed} measures, which are based on the economic output (industrial production or export).
Note that while the usual approaches rely on the UN-COMTRADE data, which only tracks physical goods, our measure of economic complexity also incorporates services.

\subsection{Measures of revealed economic complexity}

Economic complexity measures typically apply suitable algorithmic methods to infer the capability content of products and the capability endowments of territories. Such algorithms are applied to a bipartite network, which, generally speaking, connects territories to economic activities - in our case, the county-industry network $\mathbf{M^{(3)}}$. Note that, differently from the approach described in the previous sections, the information we leverage here does not regard the input of industries but their economic \textit{output}, that is, products or services. Since no external information beyond $\mathbf{M^{(3)}}$ is used in the computation, we will call it \textit{endogenous fitness}. In formulas, the Fitness of county $c$ and the Complexity of industry $i$ are given by the fixed point of the coupled equations \cite{tacchella2012new}
\begin{equation}
F_{c}^{(n)}=\sum_{i} M^{(3)}_{ci} Q_{i}^{(n-1)} \qquad
Q_{i}^{(n)}=\frac{1}{\sum_{c} M^{(3)}_{ci} \frac{1}{F_{c}^{(n-1)}}}
\label{eq:fit_endo}
\end{equation}
and follow the same line of reasoning as for Eqs. (\ref{eq:FJobs}). The equations above compute the fitness values as functions of the complexity values, and vice versa. The fitness above is our endogenous, output-based assessment of the capability endowment of counties, and is computed as a complexity-weighted diversification. The complexity is instead computed by using the strongly nonlinear argument that high-complexity industries can only be present in high-fitness counties, and, in general, ubiquity is a sign of simplicity. Following \cite{tacchella2012new}, at each iterative step, we normalize the values using the respective average values. In this case, the derived measures of economic complexity also include services.\\
Finally, following \cite{operti2018dynamics}, we compute the \textit{exogenous fitness}. In this case, we apply the Economic Fitness and Complexity algorithm to the export matrix $\mathbf{M^{(exp)}}$, which allows us to compute the complexity of products - that, however, are classified using UN-COMTRADE's Harmonized System (HS). So the next step consists in using a correspondence table \cite{pierce2012concordance} to map
HS codes to the NAICS classification used in BLS for North-American industries. The complexity values of NAICS industries are then computed using a weighted average of the HS product complexities, where the weights are given by the U.S. worldwide export values. Once computed the complexity of industries $Q_i^{(exp)}$ using the export data, we can define the \textit{exogenous fitness} as
\[
F_{c}^{Exog-Exp}=\sum_{i} M^{(3)}_{ci} Q_{i}^{(exp)} .
\]
In this case, there is no need to iterate the coupled equations: the Fitness of county $c$ is simply given by the sum of the complexities of its industries. As before, Fitness values are normalized using the mean across counties. This fitness measure is called exogenous because it relies on complexity measures computed at the country level. Note that, since the export data does not cover services but only physical goods, this issue is reflected in the exogenous Fitness computation above as well. Like the endogenous fitness, the exogenous fitness is a Revealed economic complexity measure, since the computation is based on the export or production output.

\section{Results}
\subsection{Hidden vs. Revealed Complexity of Industries: a comparison}
\begin{figure}[h]
	\centering
	\includegraphics[width=1\textwidth]{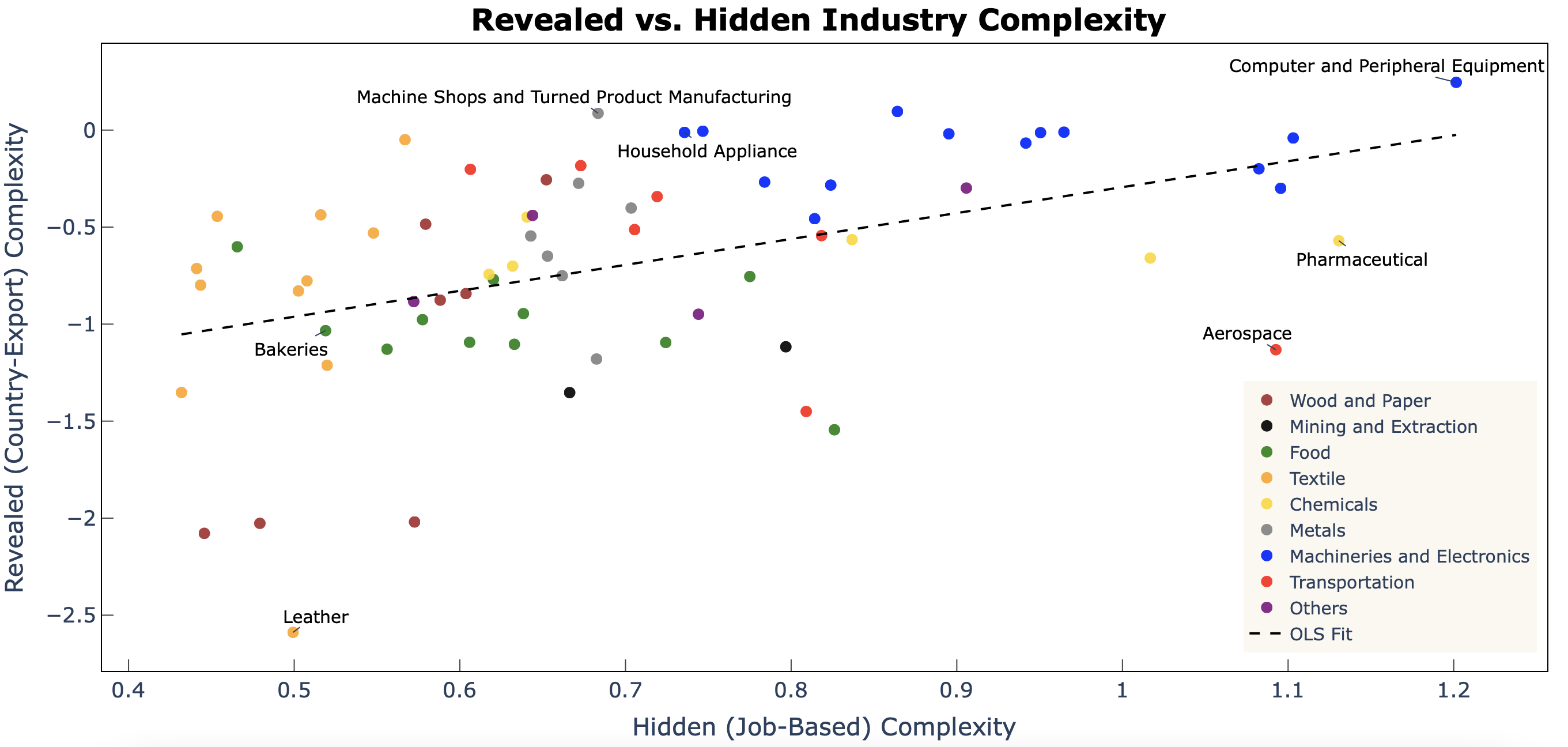}
	\caption{Comparison between the revealed complexity, based on the country export data, and our measure based on occupations. We find a general agreement between the two measures, but with notable exceptions such as Aerospace and Machine Shops. Four outliers, characterized by very low values of revealed complexity, are not shown.}
	\label{fig_compl}
\end{figure}
In Fig.\ref{fig_compl} we compare the results of two different ways to compute the complexity of industries. On the x-axis we show the measure introduced in this paper, that is, the hidden or job-based complexity. Being computed as the weighted average of the fitness of the jobs required by the industries, this is a measure based on human capabilities. On the y-axis, we report the revealed complexity, which is computed by applying the Fitness and Complexity algorithm on the country-product network \cite{tacchella2012new}. Being based on international trade, this measure leverages the industrial output as is usually done in the economic complexity literature \cite{hidalgo2021economic}. Note that, in principle, both approaches should highlight the capability content and, as such, assess the level of sophistication and technological development of industries. Indeed, by inspecting Figure \ref{fig_compl} we notice a certain agreement: Machinery and Electronics industries have a high complexity according to both measures, while Wood and Paper, Leather, and Bakeries have low complexity. However, the measures diverge in some notable examples, such as Pharmaceuticals and Aerospace, that have high job-based complexity but low revealed complexity, and Household Appliances, for which the opposite is true. Intuitively, we would attribute a high level of sophistication and technological complexity to Pharmaceutical and Aerospace, and in this respect we feel that, in these cases, the hidden complexity is providing a better assessment of the capability content of industries. On the other hand, the underestimation of Aerospace (in terms of the revealed measure) is due to the strong nonlinearity in the complexity equation, which implies that even one country re-exporting these products (such as Fiji or Bahrain re-exporting aeroplanes) leads to a lower complexity. Another consequence of this nonlinearity is the presence of very low values of revealed complexity for sectors connected to metal mining, oil and gas extraction, and petroleum derivatives (these outliers are not shown in the figure). A notable example is precisely \textit{Oil and Gas Extraction}, which is placed last in the revealed complexity ranking, at odds with the job-based complexity measure. This discrepancy underlines the operative difference between the two approaches: while the revealed complexity penalizes the capability-independent endowments of raw materials \cite{pietronero2017economic}, the hidden one rewards the complex skills required to work in this specific sector.


\subsection{Job-based complexity, labor productivity, and wages}

\begin{figure}[h!]
	\centering
	\includegraphics[width=1\textwidth]{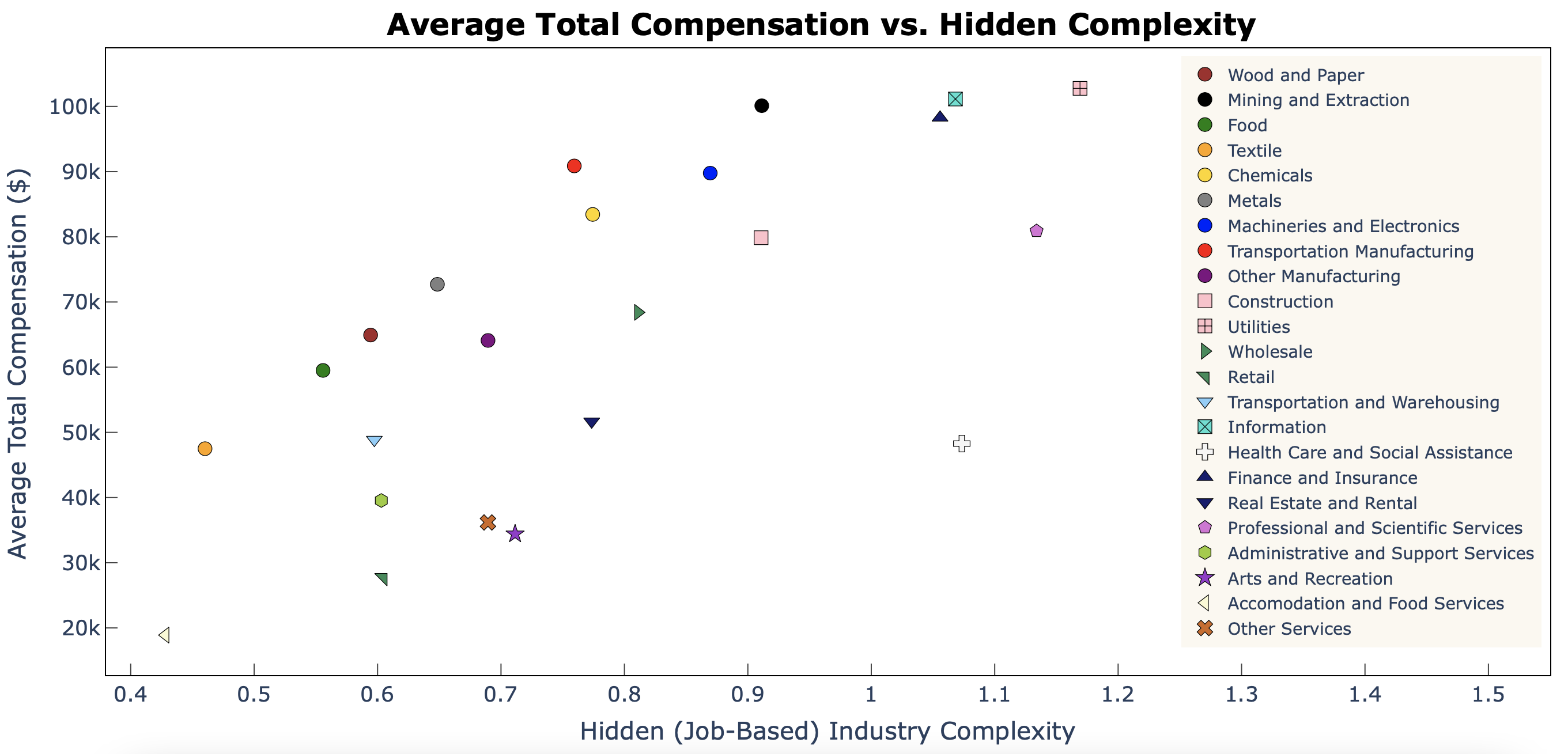}
	\caption{Positive association between job-based complexity and compensation at the sectoral level. Wage levels first increase with hidden complexity and then plateau.}
	\label{fig_wages}
\end{figure}

In this section we generalize and quantify the above intuitions by comparing the complexity measures with economic indicators. In particular, in Figure \ref{fig_wages} we show the relationship between hidden complexity and wage levels. 
Each point represents a weighted average over the industries of a specific sector, where the weights are given by the number of employees. The relationship between hidden complexity and compensation is positive for low to medium complexity, and reaches a plateau for higher levels of complexity. A notable exception is \textit{Health Care and Social Assistance}, a heterogeneous sector which includes not only specialized medical doctors but also low-skilled social workers.

\begin{table}[h]
\centering
\resizebox{0.9\textwidth}{!}{
\begin{tabular}{lccccc}
\toprule
& \textbf{[1]} & \textbf{[2]} & \textbf{[3]} & \textbf{[4]} & \textbf{[5]}\\
\midrule
$Q^{Hidden}_{17}$                          &           & 0.914***  & 0.998***  & 0.695***  & 0.914***   \\
                                           &           & (0.092)   & (0.129)   & (0.100)   & (0.091)    \\
[0.7em]
$Q^{Revealed}$                        & 0.025     &           &           &                        \\
                                           & (0.027)   &           &           &                        \\
[0.7em]
$\log(Revenues_{17})$                      & 0.328***  & 0.183***  & 0.292***  & 0.350***  & 0.183***    \\
                                           & (0.050)   & (0.026)   & (0.036)   & (0.033)   & (0.026)     \\
[0.7em]
$\log(Emp_{17})$                           & -0.281*** & -0.154*** & -0.304*** & -0.397*** & -0.154***  \\
                                           & (0.072)   & (0.033)   & (0.043)   & (0.035)   & (0.032)    \\
[0.7em]                             
$\mathrm{CR4}_{17}$                        & 0.000     & 0.002     & -0.001    &  -0.002   & 0.002     \\
                                           & (0.003)   & (0.001)   & (0.002)   &  (0.002)  & (0.001)   \\
[0.7em]                              
\textit{Service}                           &           &           &            &           & -1.315*   \\
                                           &           &           &            &           & (0.566)   \\
[0.7em]
\textit{Service} x $Q^{Hidden}_{17}$       &           &           &            &           & 0.084     \\
                                           &           &           &            &           & (0.159)   \\
[0.7em]
\textit{Service} x $\log(Revenues_{17})$   &           &           &            &           & 0.109*    \\
                                           &           &           &            &           & (0.045)   \\
[0.7em]
\textit{Service} x $\log(Emp_{17})$        &           &           &            &           & -0.150**  \\
                                           &           &           &            &           & (0.054)   \\
[0.7em]
\textit{Service} x $\mathrm{CR4}_{17}$     &           &           &            &           & -0.003    \\
                                           &           &           &            &           & (0.002)   \\
[0.7em]
\textit{Constant}                          & 6.435***  & 7.825***  & 6.510***  & 6.612***  & 7.825***  \\
                                           & (0.451)   & (0.331)   & (0.459)   & (0.449)   & (0.327)   \\
[0.7em]
\midrule
$R^2$                                      & 0.592    & 0.828   & 0.602   & 0.575  & 0.683     \\
$\mathrm{Adj.\ } R^2$                      & 0.567    & 0.818   & 0.591   & 0.567  & 0.669     \\
\text{AIC}                                 & -22.38   & -88.30  & 119.68  & 163.09 & 108.57    \\
\text{BIC}                                 & -11.00   & -76.78  & 134.59  & 180.05 & 142.51    \\
\text{F-statistic}                         & 33.597   & 120.621 & 48.064  & 60.251 & 92.341    \\
\text{p-value (F)}                         & 0.000    & 0.000   & 0.000   & 0.000  & 0.000      \\
\text{Observations}                        & 72       & 74      & 146     & 220    & 220        \\
\bottomrule
\end{tabular}
}
\caption{Linear regression model to predict the average compensation of industries. The export-based, or revealed complexity, can be computed for only 72 industries and is not statistically significant (model [1]). The job-based, or hidden complexity is always statistically significant and can be computed for goods-producing industries (model [2]), services (model [3]), and both (model [4]). Finally, in model [5] we use a dummy variable to distinguish between the two types of industries, finding that most of the difference arises from the constant term. Robust (Eicker-Huber-White) standard errors in brackets. Notation: + significant at 10\% level; * significant at 5\% level; ** significant at 1\% level; ***significant at 1‰ level.}
\label{tab_wages}
\end{table}

In order to quantify the statistical relation between these quantities, we considered four regression models that can be expressed in the form
\begin{multline}
\log(W/L)_{17,i} = \beta_0 + \beta_1Q_i + \beta_2\log(\mathrm{Emp})_{17,i} +\beta_3\log(\mathrm{Revenues})_{17,i} + \beta_4\mathrm{CR4}_{17,i} + \epsilon_i
\end{multline}
where $W/L$ is the average compensation and $Q$ is one of the complexity measures. The models share as the dependent variable the logarithm of the average total compensation of industries $W/L$. Suitable control variables are considered: revenues, number of employees, and the 4-firm Concentration Ratio CR4 (the combined market share of the 4 largest firms in each industry of each sector). 
The results of our estimations are reported in Table \ref{tab_wages}. The first model considers the revealed complexity as an explanatory variable. Note that its impact is not statistically significant. On the contrary, the job-based complexity is always significant, even if different samples are considered: goods-producing industries (model 2, similar to model 1), services (model 3), and all industries (model 4). Note how the computation of the complexity from occupations also allows its computation for services, which, however, seem to be less connected with wages. In order to investigate the possible heterogeneity of our sample, and in particular the differences between services and goods-producing industries, we evaluated a fifth model:
\begin{multline}
\log(W/L)_{17,i} = (\beta_0 + \delta_0 \cdot \mathrm{Serv}_i) + (\beta_1 + \delta_1 \cdot \mathrm{Serv}_i)Q_i + (\beta_2 + \delta_2 \cdot \mathrm{Serv}_i)\log(\mathrm{Emp})_{17,i} +\\ +(\beta_3 + \delta_3 \cdot \mathrm{Serv}_i)\log(\mathrm{Revenues})_{17,i} + (\beta_4 + \delta_4 \cdot \mathrm{Serv}_i)\mathrm{CR4}_{17,i} + \epsilon_i
\end{multline}
where the dummy variable $\mathrm{Serv}_i$ is equal to 1 if industry $i$ is a service and 0 if, instead, it is a goods-producing industry. Thus, the coefficients $\delta$ estimate the possible differences between services and goods-producing industries relative to the constant and to the coefficient associated with the independent variables. As can be seen in the last column in Table \ref{tab_wages}, we obtain quite heterogeneous estimates of the $\delta$s in terms of statistical significance. The estimate of $\delta_0$ turns out to be negative and significant at 5\%; this indicates a lower level for the average total compensation of services with respect to goods-producing industries. The estimate of $\delta_1$ turns out to be non-significant, so we cannot reject the hypothesis that the coefficient associated with Hidden Complexity is the same for goods-producing and service industries. Finally, we conducted a Chow test to assess whether the same model is able to describe both goods-producing and service industries (in this case, all $\delta$ estimates would be compatible with 0). When $\delta_0$ is included, the null hypothesis described above is rejected. When $\delta_0$ is not included, the hypothesis is only weakly rejected, indicating that most of the difference between services and goods-producing lies in the baseline compensation and so does not strongly depend on the independent variables.\\

\begin{table}[h!]
\centering
\resizebox{0.9\textwidth}{!}{
\begin{tabular}{lccccc}
\toprule
& \textbf{[1]} & \textbf{[2]} & \textbf{[3]} & \textbf{[4]} & \textbf{[5]} \\
\midrule
$Q^{Hidden}_{17}$                          &           & 0.237*    & 0.154**    & 0.207***  & 0.237*   \\
                                           &           & (0.107)   & (0.050)    & (0.040)   & (0.106)  \\
[0.7em]
$Q^{Revealed}$                        & -0.016    &           &            &           &           \\
                                           & (0.025)   &           &            &           &           \\
[0.7em]
$\log(Revenues_{17})$                      & -0.014    & -0.032    & -0.028+   & -0.045*** & -0.032   \\
                                           & (0.027)   & (0.029)   & (0.015)    & (0.013)   & (0.029)   \\
[0.7em]
$\log(Emp_{17})$                           & -0.012    & -0.003    & 0.032+    & 0.049**   & -0.003    \\
                                           & (0.037)   & (0.038)   & (0.018)    & (0.015)   & (0.037)   \\
[0.7em]                             
$\mathrm{CR4}_{17}$                        & -0.003*   & -0.003+  & 0.000      &  -0.000   & -0.003+  \\
                                           & (0.001)   & (0.001)   & (0.001)    &  (0.001)  & (0.001)   \\
[0.7em]                              
\textit{Service}                           &           &           &            &           & -0.484    \\
                                           &           &           &            &           & (0.384)   \\
[0.7em]
\textit{Service} x $Q^{Hidden}_{17}$       &           &           &            &           & -0.083    \\
                                           &           &           &            &           & (0.117)   \\
[0.7em]
\textit{Service} x $\log(Revenues_{17})$   &           &           &            &           & 0.004     \\
                                           &           &           &            &           & (0.032)   \\
[0.7em]
\textit{Service} x $\log(Emp_{17})$        &           &           &            &           & 0.036     \\
                                           &           &           &            &           & (0.042)   \\
[0.7em]
\textit{Service} x $\mathrm{CR4}_{17}$     &           &           &            &           & 0.003+    \\
                                           &           &           &            &           & (0.002)    \\
[0.7em]

\textit{Constant}                          & 0.477+   & 0.677*    & 0.193      & 0.369+   & 0.677*    \\
                                           & (0.285)   & (0.312)   & (0.227)    & (0.190)   & (0.309)     \\
[0.7em]
\midrule
$R^2$                                      & 0.165     & 0.239     & 0.055      & 0.134     & 0.191   \\
$\mathrm{Adj.\ } R^2$                      & 0.115     & 0.195     & 0.028      & 0.118     & 0.156   \\
\text{AIC}                                 & -95.11    & -103.74   & -94.64     & -178.72   & -183.57 \\
\text{BIC}                                 & -83.73    & -92.22    & -79.72     & -161.75   & -149.63 \\
\text{F-statistic}                         & 3.606     & 4.053     & 2.811      & 12.928    & 8.402   \\
\text{p-value (F)}                         & 0.010     & 0.005     & 0.028      & 0.000     & 0.000   \\
\text{Observations}                        & 72        & 74        & 146        & 220       & 220     \\
\bottomrule
\end{tabular}
}
\caption{Linear regression model to predict the labor productivity growth of industries between 2017 and 2022. All predictors are relative to the year 2017. The export-based, or revealed complexity, can be computed for only 72 industries and is not statistically significant (model [1]). The job-based, or hidden complexity is always statistically significant and can be computed for goods-producing industries (model [2]), services (model [3]), and both (model [4]). Finally, in model [5] we use a dummy variable to distinguish between the two types of industries, finding that the same model can explain both services and manufacturing. Robust (Eicker-Huber-White) standard errors in brackets. Notation: + significant at 10\% level; * significant at 5\% level; ** significant at 1\% level; ***significant at 1‰ level. }
\label{table_LPg}
\end{table}

We obtain similar results in Table \ref{table_LPg}, where the dependent variable is the Labor Productivity growth between 2017 and 2022, measured as the difference of the logarithms of gross output per worker between 2022, adjusted with PPI (Producer Price Index), and the base year 2017. The main difference with respect to the previous regression is that we cannot reject the hypothesis that goods-producing and service industries follow the same model. Note that Table \ref{table_LPg} shows the results of an out-of-sample forecast, in which we use independent variables relative to 2017 to predict a growth rate between 2017 and 2022.

\subsection{Hidden vs Revealed Fitness of counties: a comparison}
\begin{figure}[h!]
	\centering
    \includegraphics[width=0.7\textwidth]{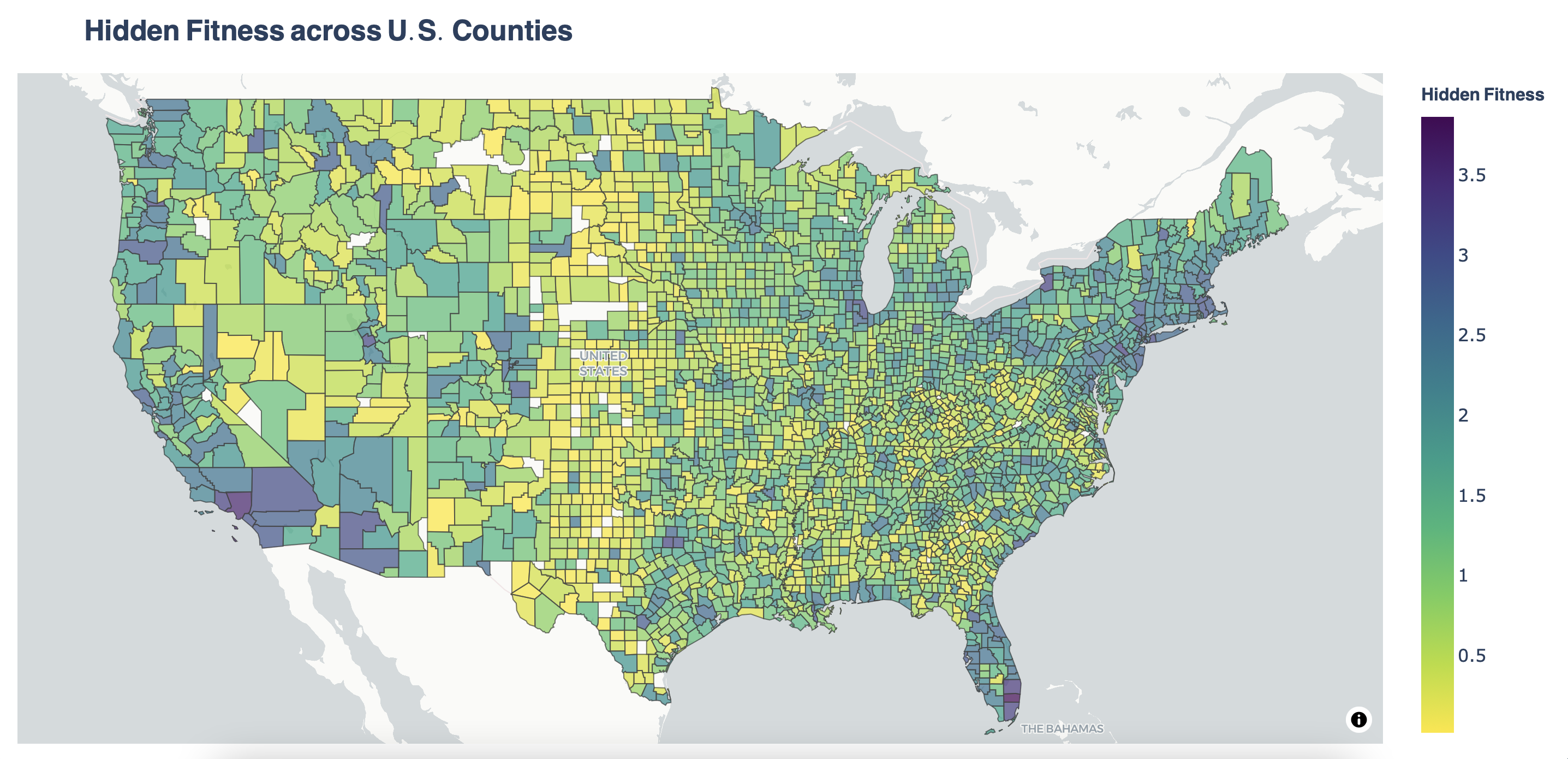}
    \includegraphics[width=0.7\textwidth]{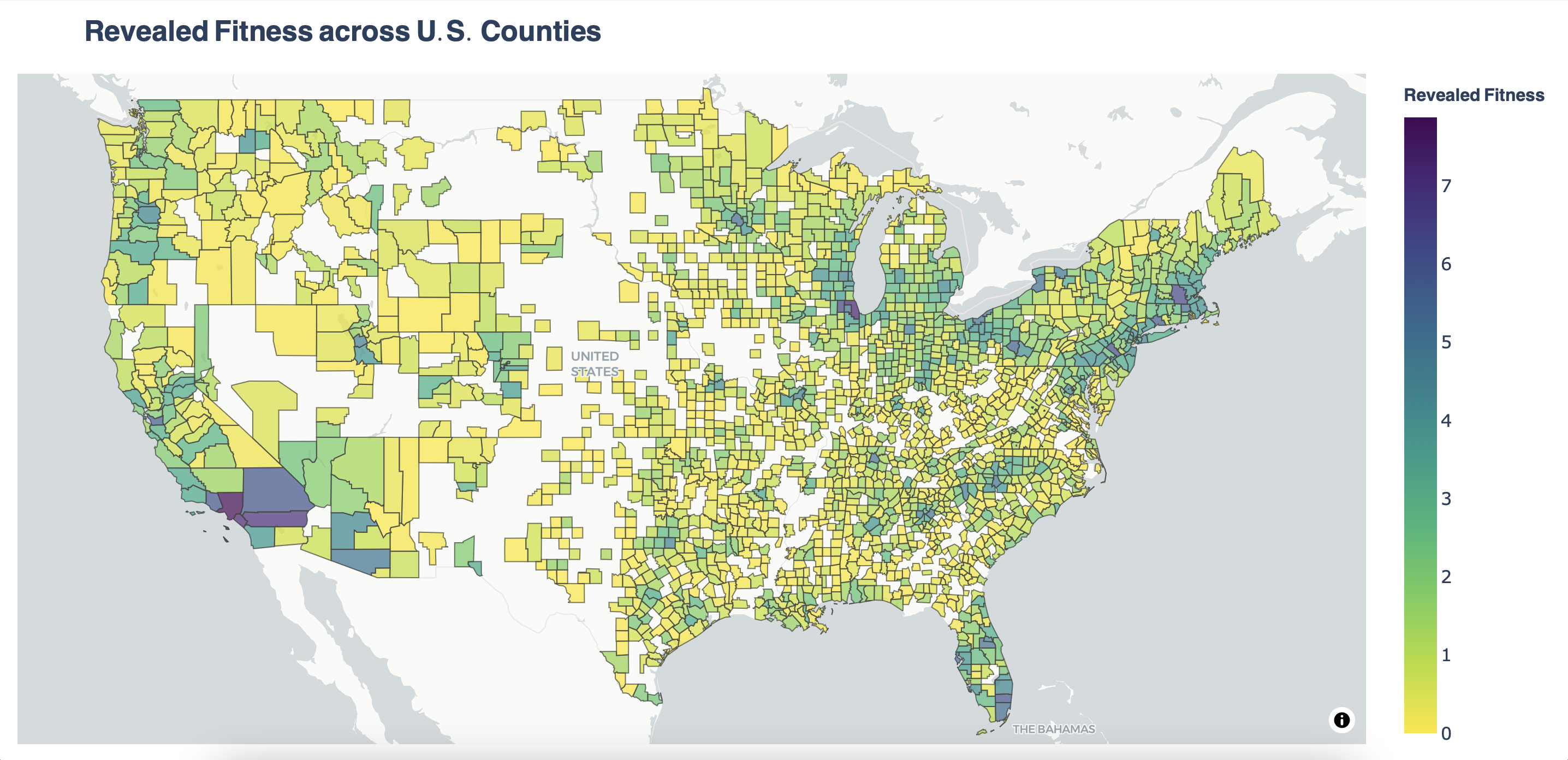}
    \includegraphics[width=0.7\textwidth]{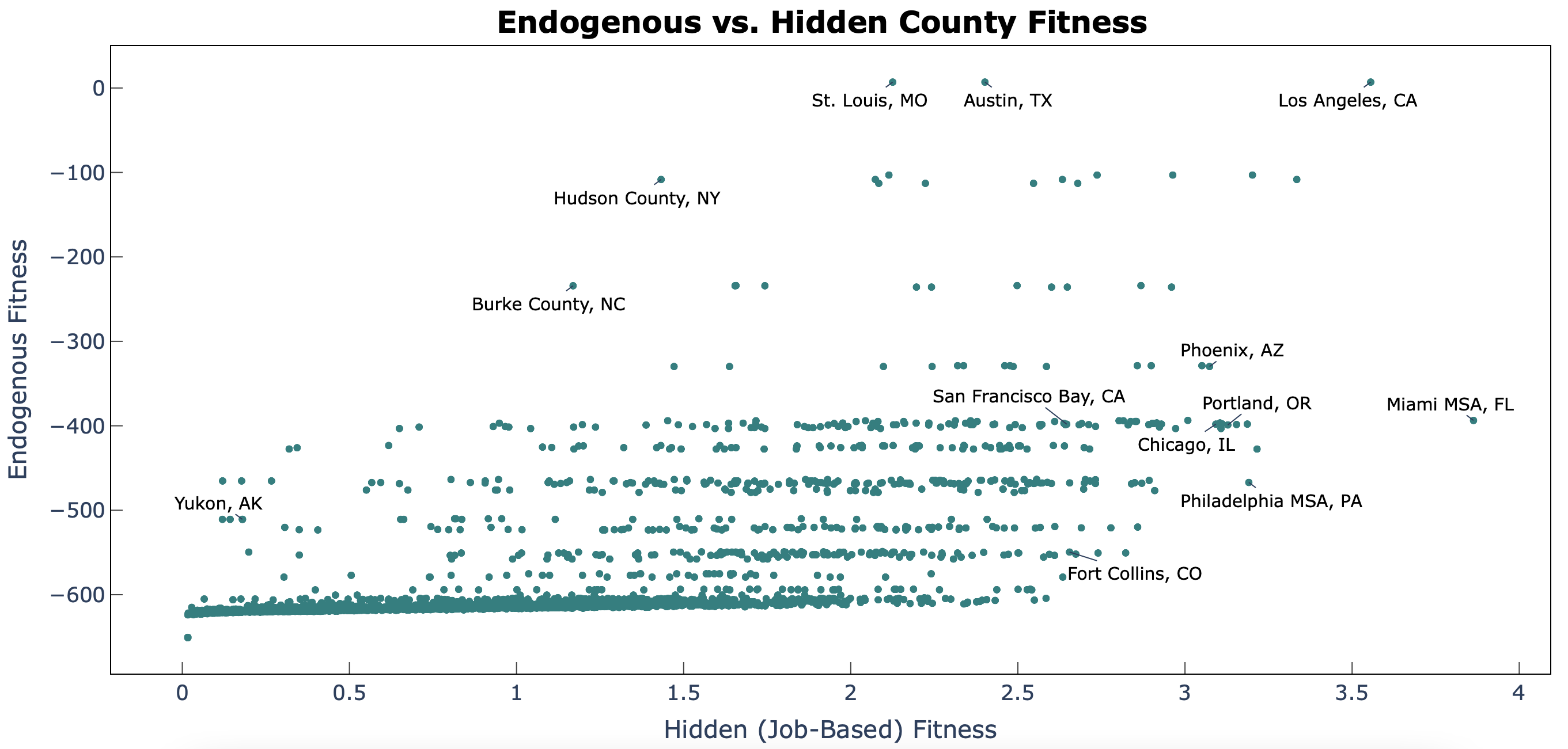}
    \caption{Top panel: job-based assessment of the economic complexity of US counties. Middle panel: exogenous fitness measure, computed using export-based complexity values. About one third of the values are not computable. Bottom panel: comparison between job-based and endogenous fitness - the latter being computed by applying the fitness algorithm to the county-industry network. The hidden fitness values are distributed more smoothly and regularly.}
	\label{fig_maps}
\end{figure}

As discussed in the previous section, the introduction of a job-based measure of industrial complexity allows for the introduction of an assessment also for service sectors, which is usually not possible since services are either unavailable or available at a very aggregated level \cite{patelli2022integrated}. In addition to this, we are now also able to compute the economic complexity of those territories that do not have a comparative advantage in goods-producing sectors. This is represented in the first two panels of Figure \ref{fig_maps}. In the top panel we represent the map of US counties, where the color is given by the hidden fitness computed as the sum of the complexity of the industries in which the county has a wage location quotient higher than 1. We are able to compute the fitness of almost all US counties. When the revealed fitness is computed, on the contrary, many counties are left without any fitness estimate, as shown in the middle panel. This is true, in particular, for those less developed counties in which almost no goods-producing industry is present (note that here we consider only nonfarm industries due to limitations in OEWS data, and that the wage location quotient is computed considering only such industries).  
As shown in the bottom panel, moreover, the hidden complexity also solves the discretization problems arising from the implementation of the endogenous fitness. In this scatter plot we show the endogenous fitness (that is, the fitness computed by applying the economic fitness and complexity algorithm to the county-industry network, see Eq.\ref{eq:fit_endo}) as a function of the job-based fitness (given by Eq.\ref{eq:fitJB}).
As a consequence of the sparsity of the county-industry network, most of the endogenous fitness values of the counties converge to very small or zero values, in line with what is expected from the investigation presented in \cite{pugliese2016convergence}. Morever, these small values accumulate in horizontal clusters and produce a multimodal distribution of fitness, which poses problems for possible linear regression analyses. This behavior is due to the presence of rare industries such as Central Banks and Tobacco Manufacturing, which get a very high level of complexity. Note that these oligopolies are in some cases a mere consequence of the privacy settings of the QCEW data and not a reflection of the economic structure of US counties. This effect is not present in the job-based fitness case, which presents a well-behaved distribution of values.

\subsection{Job-based fitness and economic growth}

In this section we analyse whether our job-based assessment of the economic complexity of counties is also able to predict their economic growth. Consistent with the empirical literature on economic growth, Table \ref{tab:gdppc} thus regresses the rate of growth of counties real GDP per capita between 2017 and 2022 on a set of independent and control variables computed for the year 2017.

\begin{table}[ht]
\centering
\resizebox{0.7\textwidth}{!}{
\begin{tabular}{lcccc}
\toprule
& \textbf{[1]} & \textbf{[2]} & \textbf{[3]} \\
\midrule
$\log({GDPpc}_{17})$          & -0.663**         & -0.653**        & -0.274          \\
                              & (0.193)          & (0.203)         & (0.276)          \\
[0.7em]
$\mathrm{divers}_{17}$        & -0.129***        & 0.006           &                  \\
                              & (0.021)          & (0.004)         &                  \\
[0.7em] 
${F}^{Job-Based}_{17}$           & 3.994***         &                 &                  \\
                              & (0.637)          &                 &                  \\
[0.7em]
${F}^{Endog}_{17}$          &                  & 0.427***        &                  \\
                              &                  & (0.078)         &                  \\
[0.7em]
$\mathrm{divers}_{17, res}$   &                  &                 & -0.539***        \\
                              &                  &                 & (0.120)          \\
[0.7em]
${F}^{Exog-Exp}_{17}$         &                  &                 & 2.691***         \\
                              &                  &                 & (0.544)          \\
\midrule
$R^2$                       & 0.075            & 0.074            & 0.095            \\
$\mathrm{Adj.\ } R^2$       & 0.058            & 0.057            & 0.071            \\
\text{AIC}                  & 14552            & 14555            & 8824             \\
\text{BIC}                  & 14570            & 14573            & 8841             \\
\text{Observations}         & 3050             & 3050             & 2079             \\
\bottomrule
\end{tabular}
}
\caption{The dependent variable is the real GDP per capita growth rate of counties. State fixed effects are included in all specifications. Cluster-robust standard errors in brackets. Notation: + significant at 10\% level; * significant at 5\% level; ** significant at 1\% level; ***significant at 1‰ level. All the measures of fitness have a positive and statistically significant effect on GDP per capita.}
\label{tab:gdppc}
\end{table}

All the specifications in Table \ref{tab:gdppc} include the lagged level of GDP per capita in order to test for the convergence hypothesis. A negative coefficient on this variable would indicate that counties which initially have a lower level of GDP per capita tend to grow faster than the counties with higher GDP, and therefore converge towards them. The specifications in columns [1] to [3] include alternative indicators of fitness and of diversification. Model [1] uses the job-based fitness computed using Eq.\ref{eq:fitJB} as a predictor, Model [2] uses endogenous fitness, and Model [3] the exogenous, export-based fitness indicator.  Inclusion of the last indicator involves a loss in the sample size of Model [3] relative to models [1] and [2]. In particular, the counties excluded from model [3] are on average less developed than the other counties. Besides the initial level of GDP per capita and the fitness indicators, the controls also comprise a diversification index, computed as the number of industries in which the counties have a wage location quotient higher than one. The presence of this variable is important in order to disentangle the effect of being active in many industries from their complexity values. The coefficient on lagged income per capita is  negative in all the specifications, although it is smaller and statistically insignificant in Model [3]. This result can be interpreted as evidence of convergence of the poorer counties towards the wealthier ones. The coefficient on fitness is positive and strongly significant in all the specifications. All the measures of fitness that we have computed are therefore positively associated with a higher rate of growth of income per capita. Interesting, however, diversification attracts a negative and statistically significant coefficient in Models [1] and [3], and is statistically insignificant in Model [2]. Diversification therefore not only does not appear to be linked to improved economic performance, but could actually be associated with a reduction in the rate of growth of output per capita.  High-complexity industries are associated with faster growth, but being active in many industries \textit{per se} does not automatically lead to growth: quite on the contrary, diversification without fitness could actually harm the growth prospects of the county.

\section{Discussion and conclusions}
Countries compete and grow thanks to their endowments, which are intrinsically heterogeneous and hard to quantify and measure. The economic complexity approach investigates the growth performance of countries and regions by assuming that their set of endowments, called capabilities, is somehow synthesized and represented by the export basket of countries. The main assumption is that a country is able to export a product if and only if it possesses all the capabilities that the specific product requires. Using this theoretical framework, a measure of the economic complexity of a product aims at quantifying the diversity and the quality of the capabilities associated with that product, and that a similar argument also applies to countries and regions. However, the capabilities are (almost) unobservable for both practical and theoretical reasons: from the lack of harmonized databases to the vagueness of their definition. As a consequence, the economic complexity approach leverages only the output information: the production and export baskets of countries. By applying suitable algorithms to the bipartite network defined by the countries and the products they export, one is able to derive an estimate of the economic complexity of these nodes, that is, a synthetic description (just one number) of the capabilities associated with a specific product or a specific country. This is what, in this paper, we call revealed complexity: an assessment of the intangible assets based on a visible output. In this paper, we reverse this perspective and quantify the economic complexity of industries and US counties starting from the capabilities, in particular, the ones embedded in the human capital. We develop a multipartite framework to connect four layers of nodes: skills, occupations, industries, and US counties. From the skill-occupation network, we algorithmically compute an assessment of the economic complexity, or fitness, of jobs. Then we compute our estimate of the complexity of industries by averaging the fitness values of the jobs active in a specific industry. Finally, the fitness of US counties is the sum of the complexity of the industries that operate in that territory. To enhance the contrast between the revealed, output-based measures of complexity, we call the input-based measures hidden complexity. This reflects our intent to investigate the hidden layer of capabilities, which connect territories to industries but are rarely analyzed. Our approach has a number of advantages. First, it allows for a natural inclusion of services, a fundamental aspect that is often neglected in the economic complexity investigations because of data unavailability. Second, it permits the computation of the fitness of those counties with a small or null manufacturing activity. Third, it solves the numerical problems of the algorithmic assessments of the economic complexity indicators, namely the presence of very small values and the emergence of multimodal distributions. Even more importantly, the hidden complexity is positively associated with local economic growth and predicts both wage levels and labor productivity growth, while the revealed complexity shows no statistical significance.

\section{Acknowledgements}
A.Z. and P.S. acknowledge the PRIN project No. 20223W2JKJ “WECARE”, CUP B53D23003880006, financed by the Italian Ministry of University and Research (MUR), Piano Nazionale Di Ripresa e Resilienza (PNRR), Missione 4 “Istruzione e Ricerca” - Componente C2 Investimento 1.1, funded by the European Union - NextGenerationEU. The authors thank Dario Sangiovanni for early investigations.
\bibliographystyle{unsrt}  
\bibliography{references}  

\end{document}